\documentclass[twocolumn]{revtex4-1}

\usepackage{amsmath}
\usepackage{graphicx}
\usepackage{amssymb}
\usepackage{color}

\newcommand{\be}{\begin{eqnarray}}
\newcommand{\ee}{\end{eqnarray}}

\newcommand{\p}{\partial}

\newcommand{\dbar}{\lower0.1ex\hbox{$\mathchar'26$}\mkern-12mu d}
 \newcommand{\kbar}{\lower0.1ex\hbox{$\mathchar'26$}\mkern-10mu k}

\newcommand{\sF}{{\mathsf F}}

\newcommand{\nn}{\nonumber}

\newcommand{\Tr}{\mathop{\rm Tr}\nolimits}
\newcommand{\diag}{\mathop{\rm diag}}
\newcommand{\cL}{{\mathcal L}}

\newcommand{\cA}{{\mathcal A}}
\newcommand{\hcA}{ \hat{{\mathcal A}}}
\newcommand{\hcB}{ \hat{{\mathcal B}}}

\newcommand{\hcF}{ \hat{{\mathcal F}}}

\newcommand{\bsF}{\boldsymbol{F} }

\newcommand{\bs}{\boldsymbol}

\newcommand{\bscF}{ \boldsymbol{{\mathcal F}}}

\newcommand{\bscA}{\boldsymbol{{\mathcal A}}}

\newcommand{\bscD}{ \boldsymbol{{\mathcal D}}}

\newcommand{\hcW}{ \hat{{\mathcal W}}}
\newcommand{\hcV}{ \hat{{\mathcal V}}}

\newcommand{\hcS}{ \hat{{\mathcal S}}}

\newcommand{\bscM}{ \boldsymbol{{\mathcal M}}}

\newcommand{\hA}{\hat{A}}
\newcommand{\hphi}{\hat{\phi}}
\newcommand{\hB}{\hat{B}}

\newcommand{\bshcF}{ \boldsymbol{\hat{\mathcal F}}}
\newcommand{\bshcA}{\boldsymbol{\hat{\mathcal A}}}
\newcommand{\bshcD}{ \boldsymbol{\hat{\mathcal D}}}
\newcommand{\bshcW}{ \boldsymbol{\hat{\mathcal W}}}

\newcommand{\bshF}{ \boldsymbol{\hat{F}}}
\newcommand{\bshA}{\boldsymbol{\hat{A}}}
\newcommand{\bshD}{ \boldsymbol{\hat{D}}}

\newcommand{\bshsF}{ \boldsymbol{\hsF} }

\newcommand{\hF}{\hat{F}}

\newcommand{\cB}{\mathcal B }
\newcommand{\cF}{{\mathcal F}}

\newcommand{\hD}{\hat{D}}

\newcommand{\hsF}{{\hat{\sF}}}
\newcommand{\mG}{{\mathrm G}}

\newcommand{\cV}{{\mathcal V}}

\newcommand{\cD}{{\mathcal D}}

\newcommand{\hcD}{\hat{{\mathcal D}}}

\newcommand{\cG}{{\mathcal G}}

\newcommand{\hcG}{\hat{{\mathcal G}}}

\newcommand{\whcG}{\widehat{{\mathcal G}}}

\newcommand{\hmJ}{\hat{\mJ}}

\newcommand{\cW}{{\mathcal W}}
\newcommand{\cS}{{\mathcal S}}

\newcommand{\bchi}{\bar{\chi} }

\newcommand{\1}{\mspace{1mu}}

\newcommand{\tGa}{\tilde{\Gamma }}
\newcommand{\hGa}{\hat{\Gamma }}

\newcommand{\hPhi}{\hat{\Phi}}

\newcommand{\mJ}{\mathrm J}
\newcommand{\mT}{\mathrm T}

\newcommand{\mM}{\mathsf M}

\newcommand{\whsF}{\widehat{\sF}}
\newcommand{\whmJ}{\widehat{\mJ}}

\newcommand{\mW}{\mathsf W}

\newcommand{\bM}{\bar{M}}

\newcommand{\mOm}{\mathsf \Omega }

\newcommand{\hmOm}{\hat{\mOm}}

\newcommand{\bshmOm}{\boldsymbol{\hmOm}}

\newcommand{\tla}{\tilde{\lambda}}
\newcommand{\hla}{\hat{\lambda}}

\newcommand{\chih}{\hat{\chi}}

\newcommand{\tchi}{\tilde{\chi}}

\newcommand{\hH}{\hat{H}}

\newcommand{\etat}{\tilde{\eta}}
\newcommand{\etab}{\bar{\eta}}

\newcommand{\etach}{\check{\eta}}

\newcommand{\ta}{\tilde{a}}
\newcommand{\tb}{\tilde{b}}

\newcommand{\hW}{\hat{W}}

\newcommand{\hC}{\hat{C}}

\newcommand{\sG}{\mathsf{G}}

\newcommand{\cC}{\mathcal{C}}

\newcommand{\hPsi}{\hat{\Psi}}

\newcommand{\chib}{\bar{\chi}}

\begin{document}




\def\intdk{\int\frac{d^4k}{(2\pi)^4}}
\def\sla{\hspace{-0.22cm}\slash}
\hfill


\title{A general theory of the standard model \\ and the revelation of the dark side of the universe }

\author{Yue-Liang Wu}\email{ylwu@itp.ac.cn, ylwu@ucas.ac.cn}

\affiliation{
$^1$ Institute of Theoretical Physics, Chinese Academy of Sciences, Beijing 100190, China \\
$^2$ International Centre for Theoretical Physics Asia-Pacific  (ICTP-AP), Beijing 100190, China \\
$^3$ Taiji Laboratory for Gravitational Wave Universe (Beijing/Hangzhou), University of Chinese Academy of Sciences(UCAS), Beijing 100049, China \\
$^4$ School of Fundamental Physics and Mathematical Sciences, Hangzhou Institute for Advanced Study, University of Chinese Academy of Sciences (UCAS), Hangzhou 310024, China }



\maketitle




The discovery of the Higgs boson\cite{higgs2012A,higgs2012C} marked the fulfillment of a key prediction within the standard model (SM)\cite{EW1,EW2,EW3,QCD1,QCD2}, which describes electroweak and strong interactions through the gauge symmetry $\mG_{SM}$ = U$_Y$(1)$\times$ SU$_L$(2)$\times$ SU$_C$(3). This model is firmly established in quantum field theory (QFT) within Minkowski spacetime and extensively supported by experimental verification. Meanwhile, the direct detection of gravitational waves\cite{GW} has provided crucial confirmation for the Einstein's general theory of relativity (GR) \cite{GR}, which is characterized by the dynamics of Riemannian geometry in curved spacetime based on the general symmetry GL(1,3,R).

However, reconciling GR with QFT remains a profound challenge due to fundamental differences in their theoretical frameworks. It is widely believed that the laws of nature stem from the intrinsic properties of quantum fields as elementary particles, rather than from the external geometric properties of spacetime coordinates. This perspective necessitates a clear distinction between intrinsic symmetries, governed by particle quantum numbers, and external symmetries, which describe particle motion in Minkowski spacetime. This conceptual distinction motivates the introduction of a biframe spacetime with a fiber bundle structure, where Minkowski spacetime serves as the base and spin-related intrinsic gravigauge spacetime acts as the fiber. Such a structure enables the description of gravitational interactions within the QFT framework through spin-related gauge symmetries\cite{GQFT1,GQFT2,HUFT1,HUFT2,FHUFT}, which provides a theoretical framework of gravitational quantum field theory (GQFT). Previous studies have focused on the gravidynamics\cite{GQFT1,GQFT2} and hyperunified field theory\cite{HUFT1,HUFT2,FHUFT}, with recent research exploring novel effects\cite{GQFTN} within GQFT. Where the gravigauge field as the fundamental gravitational field emerges as a bi-covariant vector field and behaves as a Goldstone-type boson in bi-frame spacetime.

In this paper, we aim to develop a comprehensive standard model that provides a unified description of all fundamental forces, integrates the two established standard models in particle physics and cosmology, and sheds light on the nature of gravity as well as the mysteries of the dark side of the universe. This proposed general theory of the standard model (GSM) is grounded in the principle of gauge invariance and incorporates the following extended internal gauge symmetries inherent in the SM:
\be \label{SGSM1}
& & \mG_{GSM} = WS_c(1,3)\times SG(1) \times \mG_{SM} , \nn \\
& & WS_c(1,3) = SP(1,3)\rtimes W^{1,3} \rtimes SP_c(1,1) ,
\ee
where WS$_c$(1,3) is referred to as conformal inhomogeneous spin gauge symmetry, with its subgroups SP(1,3), W$^{1,3}$, and SP$_c$(1,1) corresponding to spin, chirality boost-spin, and chiral conformal-spin gauge symmetries, respectively. SG(1) denotes scaling gauge symmetry independent of chirality. 

It enables to show that GSM brings rich new forces: spin, chirality boost-spin, chiral conformal-spin and scaling gauge forces. In particular, the gravitational force is mediated by gravigauge field, identified as massless graviton with five independent polarizations\cite{GQFTN}. The chirality boost-spin gauge field associated to the translation-like subgroup symmetry of WS(1,3) becomes a massive gauge field, termed the dark graviton, acts as dark matter candidate, interacting with leptons and quarks via heavy spin gauge boson. Moreover, the canonical scalar field serves as sources of primordial energy and dark energy, driving the early inflation and current accelerated expansion of the universe.

Leptons and quarks in the SM are Weyl fermions due to maximal parity violation\cite{PV}. However, in their electromagnetic interactions, their left-handed and right-handed components are treated equivalently to behave as a Dirac spinor. Namely, a Dirac spinor can always be expressed symmetrically as a superposition of left-handed and right-handed spinors via chiral spinor representation in eight-dimensional Hilbert space. Therefore, it is useful to adopt a conventional chiral spinor representation for leptons and quarks to reformulate the SM, 
\be \label{NC}
& & \Psi_{-}^{i} \equiv \binom{\Psi_{L}^{i}}{\Psi_{R}^{i}} \quad \mbox{or} \quad  \Psi_{+}^{i}\equiv \binom{\Psi_{R}^{i}}{\Psi_{L}^{i}}; \;\;  \Psi_{\mp}^i \equiv l_{\mp}^i, q_{\mp}^i ,\nn \\
& &  \Psi_{L,R}^{i} = l_{L,R}^{i} = \binom{\nu_{L,R}^{i}}{e_{L,R}^{i }} ,  \; \Psi_{L,R}^{i}= q_{L,R}^{i }= \binom{u_{L,R}^{i}}{d_{L,R}^{i}}  , 
\ee
where $\Psi_{\mp}^i$ represent two equivalent chirality representations of leptons and quarks, and $\Psi_{L,R}^{i}$ denote the doublets of left-handed and right-handed leptons and quarks with $\psi_{L,R}^{i} \equiv (\nu^{i}, e^{i}, u^{i}, d^{i})_{L,R}$ ($i=1,2,3$) labeling the three families of left-handed and right-handed neutrinos, charged leptons, up-type quarks, and down-type quarks. Note that the three families in SM remain a puzzle for further exploration, the right-handed neutrinos are treated equally to left-handed neutrinos in our present consideration as neutrinos are thought to be massive due to oscillations, which is in fact beyond the SM. Although the right-handed leptons/quarks are expressed formally as doublets, their electroweak gauge interactions remain to be considered as the same as in SM, i.e, SU$_L$(2)$\times$U$_Y$(1). Namely, we just reformulate the SM in light of the left-right symmetric chiral spinor representations of leptons/quarks rather than left-right symmetric gauge symmetries. Here, U$_Y$(1) is actually a subgroup of SU$_R$(2)$\times$U$_{B-L}$(1). It should be interesting to extend SM to have left-right symmetric gauge interactions, which is beyond the scope of present paper.    

The fundamental laws of nature should be independent of the choice between the two equivalent chirality representations, $\Psi_{-}^i$ and $\Psi_{+}^i$, which leads to a principle of chiral duality invariance. Namely, the theory should be invariant under exchange between negative and positive chirality, $ ``-" \leftrightarrow ``+"$, which exhibits a $Z_2$ discrete symmetry in the theory. Based on this principle and gauge invariance principle, we are able to build the action of GSM in a spin-related intrinsic gravigauge spacetime for all involved basic fields:
\be \label{GSMaction1}
\cS_{GSM}  & = & \int [\dbar \zeta ]  \cL_{GSM}( \Psi_{\mp}^i,  \hB_{c}, \hW_{c}^{i}, \hA_{c}^{\alpha}, \hat{H}, \nn \\
& & \hcB_{c}, \hcW_{c}^{a}, \hcA_{c}^{ab}, \hcW_{c}, \hat{\Phi}_{\kappa}, \hat{\phi}_e,\hat{\phi}_w, \bs{\zeta}^{a}, \bs{\zeta} ), 
\ee 
with $\cL_{GSM}$ the Lagrangian of GSM. Here, $\hB_{c}$, $\hW_{c}^{i}$, and $\hA_{c}^{\alpha}$ represents electroweak and strong gauge fields relating to the gauge symmetries U$_Y$(1)$\times$SU$_L$(2)$\times$SU$_C$(3) and $\hat{H}$ denotes Higgs field in SM. $\hcB_{c}$, $\hcW_{c}^{a}$ and $\hcA_{c}^{ab}$ correspond to chiral conformal-spin, chirality boost-spin and spin gauge fields associated to the conformal inhomogeneous gauge symmetry WS$_c$(1,3), and $\hcW_{c}$ stands for Weyl scaling gauge field due to scaling gauge symmetry. $\hPhi_{\kappa}$, $\hphi_{e}$ and $\hphi_w$ are three singlet scalar fields introduced to preserve scaling symmetries. $\bs{\zeta}$ and $\bs{\zeta}^{a}$ are  the scaling and vector fields, respectively, in gravigauge spacetime, which are required to construct the chiral conformal-spin and chirality boost-spin gauge invariant action.

The explicit form of the above action and relevant definitions concerning covariant derivatives and all basic fields as well as group generators of gauge symmetries $\Sigma_{Y \mp}^{(\Psi)}$, $\Sigma_{L \mp}^i$, $T_{\alpha}^{(\Psi)}$, $\Sigma_{\mp}$, $\Sigma_{a \mp}$, $\Sigma_{ab}/2$ and $\Gamma$-matrices are presented in Section A of Supplementary material. 

The action of GSM involves seven parameters including two gauge couplings ($g_4$, $g_w$) and five scalar couplings ($\beta_{\kappa}$, $\beta_w$, $\beta_e$, $\gamma_{\kappa}$ and $\gamma_w$) in addition to SM. The Yukawa couplings $\hla^{\Psi}_{ij}= (\hla^{\Psi}_{ij})^{\dagger}$ and $\tla^{\Psi}_{ij}=(\tla^{\Psi}_{ij})^{\dagger}$ ($\Psi = l,q$) are all hermitian matrices, which bring on hermitian mass matrices for neutrinos, charged leptons, up-type and down-type quarks via the combinations: $\lambda^{\nu,e} \equiv \tla^{l} \pm \hla^{l}$, $\lambda^{u,d} \equiv \tla^{q} \pm \hla^{q}$. This differs from those in SM, where neutrinos are massless and all Yukawa coupling matrices are in general arbitrary complex matrices rather than hermitian matrices. Obviously, each complex Yukawa coupling matrix in SM has 18 arbitrary parameters, which is diagonalized via bi-unitary matrices (left-handed and right-handed unitary matrices), while each hermitian Yukawa coupling matrix contains only 9 arbitrary parameters, which is diagonalized by a single unitary matrix.  

In the action of GSM, we have introduced spin-related derivative operator $\eth_c$ and displacement vector $\dbar \zeta^{c}$. Unlike the ordinary derivative operator $\partial_{\mu} \equiv \frac{\partial}{\partial x^{\mu}}$ and displacement vector $dx^{\mu}$ in coordinate spacetime, $\eth_c$ satisfies a non-commutative relation: $ [ \eth_c ,\; \eth_d] = \hsF_{cd}^a\, \eth_a $ with  $\hsF_{cd}^a$ acting as group structure factor, generating non-Abelian group algebra. 

In the external coordinate spacetime, the dual bases ${\partial_{\mu}}$ and ${dx^{\mu}}$ with inner product $\langle dx^{\mu}, \partial_{\nu} \rangle = \eta_{\nu}^{\mu}$ span the tangent and cotangent spacetimes $\mT_{\mM}$ and $\mT^{\ast}_{\mM}$. Analogously, ${\eth_a}$ and ${\dbar \zeta^{a}}$ are dual bases with inner product $\langle \dbar \zeta^c, \eth_d \rangle = \eta_d^c$, forming the spin-related dual spacetimes $\mT_{\sG}$ and $\mT^{\ast}_{\sG}$. This intrinsic spacetime is referred to as gravigauge spacetime, which produces a non-commutative geometry described by the group structure factor $\hsF_{cd}^a$.

It can be proved that the action of GSM possesses the whole gauge symmetry $\mG_{GSM}$.  Under the spin gauge transformations, $\Psi_{\mp}^{i} \to S(\varpi^{ab}) \Psi_{\mp}^{i}$ with $S(\varpi^{ab}) = e^{i\varpi^{ab}\Sigma_{ab}/2 }\in$ SP(1,3), the spin gauge field $\hcA_{c}\equiv \hcA_{c}^{\; ab}\Sigma_{ab}/2$ transforms as non-Abelian gauge field, $\hcA_{c} \to \hcA'_{c} =$ $S(\varpi^{ab}) i\eth_{c} S^{-1}(\varpi^{ab})$ $+ S(\varpi^{ab}) \hcA_{c}S^{-1}(\varpi^{ab})$, while all other gauge fields $\hcA_{c}^{A}$ transform as vector fields. For chirality boost-spin gauge transformations, $S_{\mp}(\varpi^{a}) = e^{i\varpi^{a}\Sigma_{\mp a}} \in W^{1,3}$, the chirality boost-spin gauge field $\hcW_{c}^{(\mp)} \equiv \hcW_{c}^{\; a} \Sigma_{\mp a}$ transforms as: $\hcW_{c}^{(\mp)} \to \hcW_{c}^{'(\mp)} = (\hcW_{c}^{\; a} + \hcD_{c}\varpi^{a})\Sigma_{\mp a}$, with $\hcD_{c}\varpi^{a} \equiv (\eth_c +\hcB_{c}) \varpi^{a} + \hcA_{c\, b}^{a}\varpi^{b}$. For chiral conformal-spin gauge transformations,  $S_{\mp}(\varpi) = e^{i\varpi \Sigma_{\mp}} \in SP_c(1,1)$, the chiral conformal-spin gauge field $\hcB_{c}^{(\mp)}\equiv \hcB_{c}\Sigma_{\mp}$ transforms as scaling-type gauge field, $\hcB_{c}^{(\mp)}\to  \hcB_{c}^{'(\mp)} =   (\hcB_{c} - \eth_{c}\varpi ) \Sigma_{\mp}$,  and $\hcW_{c}^{(\mp)}$ transforms as, $\hcW_{c}^{(\mp)} \to \hcW_{c}^{'(\mp)} = e^{-\varpi }\hcW_{c}^{\; a} \Sigma_{\mp a}$, which defines a negative chiral conformal-spin charge $C_c=-1$. Under the scaling gauge transformations, $S(\xi) \in SG(1)$, the scaling gauge field $\hcW_{c}$ transforms as: $\hcW_{c}\to \hcW_{c}' = \hcW_{c} - \eth_{c}\ln \xi $. It can be verified that the spinor field $\Psi_{s}$ carries positive charge $C_c=1/2$ and local scaling charge $\hC_s = 3/2$. The scalar field $\hPhi_{\mp}$ has local scaling charge $\hC_s=1$, while the spin-related derivative operator $\eth_c$ and displacement vector $\dbar \zeta^{c}$ have charges $\hC_s=1$ and $\hC_s=-1$, respectively. To maintain scaling gauge symmetry, three singlet scalar fields, $\hPhi_{\kappa}$, $\hphi_{e}$, $\hphi_w$, are introduced with positive charge $\hC_{s} = 1$. To preserve gauge symmetries $W^{1,3}$ and $SP_c(1,1)$, the vector field $\bs{\zeta}^{a}$ and scaling field $\bs{\zeta}$ are presented, both carrying negative charge $C_c= -1$. $\bs{\zeta}^{a}$ gets a translational transformation in gravigauge spacetime, $\bs{\zeta}^{a}\to \bs{\zeta}^{' a}= \bs{\zeta}^{a} + \varpi^{a}$, and $\hphi_w$ carries negative charge $C_c= -1$. $\hcV_{S}( \hPhi/\hPhi_{\kappa}, \bs{\zeta} \hphi_w/\hPhi_{\kappa}, \hphi_{e}/\hPhi_{\kappa})$ denotes an invariant scalar potential. 

The action of GSM also possesses the chiral duality symmetry under the operation $\cC_d$: $\hPsi_{\mp}^{c_d}  \equiv C_d \hPsi_{\mp} = \hPsi_{\pm}$, $\bshcD_{c}^{(\Psi \mp)c_d}\equiv C_{d}\bshcD_{c}^{(\Psi \mp)} C_{d}^{-1}  \equiv \bshcD_{c}^{(\Psi \pm)}$ and  $\Phi_{\mp}^{c_d}\equiv C_{d}\Phi_{\mp} C_{d}^{-1}  \equiv \Phi_{\pm}$. Here $C_d = \sigma_1\otimes \sigma_0 \otimes \sigma_0\otimes \sigma_0$ with $C_d^2 = 1 $ defines explicitly a permutation operation matrix for the discrete chiral duality symmetry. 

Gauge invariance allows for specific gauge fixing conditions: $\bs{\zeta}^{' a}= \bs{\zeta}^a + \varpi^a = 0$ and $\bs{\zeta}' = e^{-\varpi} \bs{\zeta} = 1$. Thus, we can always choose a unitary conformal-boost gauge basis by setting $\bs{\zeta}=1$ and $\bs{\zeta}^a= 0$, which leads to the replacements: $\bshcW_{c}^{(\mp)} \to \hcW_{c}^{(\mp)}$ and $\bshcF_{cd}^{(W\mp)} \to \hcF_{cd}^{(W\mp)}$, as shown from the definitions in Section A of Supplementary material. 

Notably, $\bshsF_{cd}^a$ has no dynamic term in Eq.(\ref{GSMaction1}) and acts as an auxiliary field,  leading to the following constraint:
\be \label{GraviE}
\bscM_{cda}^{\, c'd'a'} \bshsF_{c'd' a'}=\widehat{\bscF}_{cda}\;\; \mbox{or} \;\; \bshsF_{cd a}  = \widehat{\bscM}_{cda}^{\, c'd'a'} \widehat{\bscF}_{c'd'a'},
\ee
which facilitates determining $\bshsF_{cd}^a$. It illustrates that the gravitational effects emerging from the non-commutative nature of gravigauge spacetime are intricately governed by the collective dynamics of all fundamental gauge fields. We may refer to above equation as gravitization equation. The explicit expressions for $\bscM_{cda}^{\; c'd'a'}$ and $\widehat{\bscF}_{cda}$ are presented in Section B of Supplementary material. 
 
To elucidate further, the spin-related derivative and displacement  $\eth_a$ and $\dbar \zeta^{a}$ are defined as,  $\eth_{a}  \equiv  \chih_{a}^{\; \mu}(x)\p_{\mu}$ and $\dbar\zeta^{a} \equiv  \chi_{\mu}^{\; a}(x) dx^{\mu} $, with $\p_{\mu}$ and $dx^{\mu}$ being the derivative operator and displacement vector in coordinate spacetime, respectively. Here $\chi_{\mu}^{\; a}(x)$ and $\chih_{a}^{\; \mu}(x)$ are dual vector fields satisfying the dual conditions, $\chi_{\mu}^{\; a}(x)  \eta_{a}^{\; b} \chih_{b}^{\; \nu}(x) = \eta_{\mu}^{\; \nu} $ and 
$\chih_{b}^{\; \nu}(x) \eta_{\nu}^{\; \mu} \chi_{\mu}^{\; a}(x)  = \eta_{b}^{\; a}$, which transform as bi-covariant vector fields under both spin gauge symmetry SP(1,3) in gravigauge spacetime and global Lorentz symmetry SO(1,3) in Minkowski spacetime. Thus, $\hsF_{ab}^c$ takes an explicit form, $\hsF_{cd}^a \equiv - \chih_{c}^{\; \mu} \chih_{d}^{\; \nu} \sF_{\mu\nu}^{a}$ with $\sF_{\mu\nu}^{a} \equiv \p_{\mu}\chi_{\nu}^{\; a}(x) - \p_{\nu}\chi_{\mu}^{\; a}(x)$, which defines a field strength by regarding $\chi_{\mu}^{\; a}(x)$ as gauge-type vector field. For convienence, $\chi_{\mu}^{\; a}(x)$ is referred to as gravigauge field as it reveals gravitational interaction. It is evident that the gravigauge spacetime characterized by gravigauge field emerges as local flat non-commutative spacetime.

The dual gravigauge fields $\chi_{\mu}^{\; a}(x)$ and $\chih_{a}^{\; \mu}(x)$ allow the projection of internal vectors in gravigauge spacetime into external vectors in coordinate spacetime, $i\bscD_{\mu} \equiv \chi_{\mu}^{\; c} i\bshcD_{c}$, $\bscA_{\mu} \equiv \chi_{\mu}^{\; c}\bshcA_{c}$ and $\bscF_{\mu\nu} \equiv \chi_{\mu}^{\; c} \chi_{\nu}^{\; d} \bshcF_{cd}$. Here $\bscA_{\mu}$ and $\bscF_{\mu\nu}$ represent all relevant gauge fields and the corresponding gauge covariant field strengths.

The scaling gauge invariance allows to choose a fundamental mass scaling gauge basis by setting $\hPhi_{\kappa}(x) = \bM_{\kappa}$. This leads to the simplifications and replacements: $\hD_{c}\hPhi_{\kappa} \to \bM_{\kappa} \hcW_{c}$, $\cS_{c} =0$, ($\bshsF_{cd}^{a}$, $\bshmOm_{c}^{ab}$, $\hphi_w$, $\hphi_{e}$,  $\hH$) $\to (\hsF_{cd}^{a}$, $\hmOm_{c}^{ab}$, $\phi_w$, $\phi_{e} $, $H$), and $\hPhi_{\kappa}^4\, \hcV_{S}(\hPhi/\hPhi_{\kappa}, \hphi_w/\hPhi_{\kappa}, \hphi_{e}/\hPhi_{\kappa})\to \cV_{S}(H, \phi_w, \phi_{e})$. In the unitary conformal-boost gauge basis and fundamental mass scaling gauge basis, by utilizing $\chi_{\mu}^{\; a}(x)$ and $\chih_{a}^{\; \mu}(x)$ as projecting bi-covariant vector fields, we can reformulate the action in Eq.(\ref{GSMaction1}), whose explicit form is provided in Section A of Supplementary material, into the following formalism within the framework of GQFT based on Minkowski spacetime: 
\be \label{GSMaction2}
& & \cS_{GSM}  \equiv  \int [d x]\chi \cL_{GSM}(  \Psi_{L,R}^i,  B_{\mu}, W_{\mu}^{i}, A_{\mu}^{\alpha}, H, \nn \\
& & \qquad \; \cB_{\mu}, \cW_{\mu}^{a}, \cA_{\mu}^{ab}, \cW_{\mu}, \phi_e,\phi_w ) = \int [d x]\chi \lbrace \sum_{\Psi=l,q} \nn \\
& & \qquad  \frac{1}{2}  ( \bar{\Psi}_{L}^{i} \gamma^{a}  i \chih_{a}^{\; \mu} \cD_{\mu}^{(\Psi_L)} \Psi_{L}^{i}  +\bar{\Psi}_{R}^{i} \gamma^{a}  i \chih_{a}^{\; \mu} \cD_{\mu}^{(\Psi_R)} \Psi_{R}^{i}  \nn \\
& & \qquad + \bar{\Psi}_{L}^{i} H \lambda^{\Psi^{(-)}}_{ij} \Psi_{R}^{(-)j} +  \bar{\Psi}_{L}^{i} \tilde{H} \lambda^{\Psi^{(+)}}_{ij} \Psi_{R}^{(+)j}  + H.c. ) \nn \\
& & \qquad -  \frac{1}{4} \chih^{\mu\mu'}\chih^{\nu\nu'} ( \sum \bscF_{\mu\nu}^{A} \bscF_{\mu'\nu'}^{A}  - \cF_{\mu\nu}^{a} \cF_{\mu'\nu' a} )  \nn \\
& & \qquad + \frac{1}{4} \bM_{\kappa}^2\tchi^{\mu\nu\mu'\nu'}_{aa'}\sF_{\mu\nu}^{a}\sF_{\mu'\nu' }^{a'} + \frac{1}{2}  \beta_{\kappa}^2 g_w^2 \bM_{\kappa}^2  \chih^{\mu\nu}\cW_{\mu}\cW_{\nu} 
 \nn \\
 & & \qquad + \frac{1}{4} (\gamma_{\kappa}^2\bM_{\kappa}^2 + \beta_e \phi_{e}^2 ) \bchi^{\mu\nu\mu' \nu'}_{aa'}\cG_{\mu\nu}^{a}\cG_{\mu'\nu'}^{a'} \nn \\
& & \qquad + \frac{1}{2}  g_4^2  \phi_w^2 \chih^{\mu\nu} (\gamma_w^2 \cB_{\mu} \cB_{\nu} -\beta_{w}^2 \cW_{\mu}^{a}  \cW_{\nu a}  ) \nn \\
& & \qquad  + \frac{1}{2} \chih^{\mu\nu}  (\cD_{\mu}\phi_{w} \cD_{\nu}\phi_{w} + \cD_{\mu}\phi_{e} \cD_{\nu}\phi_{e} ) \nn \\
& & \qquad +  \chih^{\mu\nu} (\cD_{\mu}H)^{\dagger} \cD_{\nu}H  - \cV_{S}(H, \phi_w, \phi_{e})  \rbrace ,
\ee
with $\bscF_{\mu\nu}^{A} \equiv (F_{\mu\nu}, F_{\mu\nu}^{i}, F_{\mu\nu}^{\alpha}, \cF_{\mu\nu}, \cF_{\mu\nu}^{ab}, \cW_{\mu\nu})$ the well defined gauge field strengths in the summation, and $\Psi_R^{(\pm)}\equiv (l^{(+)}, l^{(-)})_R \equiv (\nu, e)_{R}, (q^{(+)}, q^{(-)})_R \equiv (u, d)_{R}$. Note that the gauge fields are rescaled as $\cA_{\mu}^{A} \to g_A \cA_{\mu}^{A}$ with $g_A$ the corresponding gauge couplings. The relation between two integral measures is taken, $[\dbar \zeta]\equiv [dx]\, \chi$ with $\chi = \det \chi_{\mu}^{\; a} $. The above action concerns definitions for the covariant derivatives, $i\cD_{\mu}^{(l_{L,R}; q_{L,R})}  \equiv i D_{\mu}^{(l_{L,R}; q_{L,R})} + g_4 \cA_{\mu}^{ab}\Sigma_{ab}/2$, $D_{\mu}\phi_{e} \equiv (\p_{\mu} + g_w \cW_{\mu} ) \phi_{e}$ and $\cD_{\mu}\phi_{w} \equiv (\p_{\mu} + g_w \cW_{\mu} ) \phi_w$. Here $D_{\mu}^{(l_{L,R}; q_{L,R})} $ and $D_{\mu}H$ ($ \tilde{H} \equiv  i \sigma_2 H^{\ast}$) are covariant derivatives with the same forms as those in SM. The composed tensor $\chi_{\mu\nu} \equiv \chi^{\;a}_{\mu} \chi^{\; b}_{\nu} \eta_{ab}$ defines the gravimetric field with its dual tensor $\chih^{\mu\nu} \equiv \chih_{a}^{\; \mu} \chih_{b}^{\; \nu} \eta^{ab}$.

It can be verified that the above action, after imposing three gauge fixing conditions, possesses a joint symmetry, $\widetilde{G}_{GSM} = SC(1)\times PO(1,3) \Join SP(1,3)\times \mG_{SM}$. Here SC(1) and PO(1,3)$\equiv$SO(1,3)$\times P^{1,3}$ denote the usual global scaling symmetry and inhomogeneous Lorentz (or Poincar\'e) symmetry in Minkowski spacetime, acting as a base spacetime, and SP(1,3)$\times \mG_{SM}$ represent internal gauge symmetries in spin-related gravigauge spacetime, functioning as a fiber. Mathematically, it forms a bi-frame spacetime with a fiber bundle structure.

The action in Eq.(\ref{GSMaction2}) represents the GSM within the framework of GQFT. This formalism enables to derive the equation of motion for the gravigauge field $\chi_{\mu}^{\; a}$ with conserved current as follows:
\be  \label{GaugeGE}
 \p_{\nu} \whsF^{\mu\nu }_{a}  = \whmJ_{a}^{\; \mu} , \quad \p_{\mu} \whmJ_{a}^{\; \mu} = 0, 
\ee
with the field strength, $\whsF^{\mu\nu }_{a} \equiv \chi \1 \tchi^{[\mu\nu]\mu'\nu'}_{a a'}  \sF_{\mu'\nu' }^{a'}$, and tensor $\tchi_{aa'}^{\mu\nu \mu'\nu'} \equiv \chih_{c}^{\;\, \mu}\chih_{d}^{\;\, \nu} \chih_{c'}^{\;\, \mu'} \chih_{d'}^{\;\, \nu'}  \etat^{c d c' d'}_{a a'}$. The current $\whmJ_{a}^{\; \mu}$ is conserved due to antisymmetric property of field strength, $\whsF^{\mu\nu }_{a} = - \whsF^{\nu\mu }_{a}$, i.e., $\p_{\mu} \whmJ_{a}^{\; \mu} = \p_{\mu}\p_{\nu} \whsF^{\mu\nu }_{a} = - \p_{\nu}\p_{\mu} \whsF^{\nu\mu }_{a} = 0$. 

The above equation of motion for the gravigauge field provides a general description on gravidynamics in GQFT, referred to as gauge-type gravitational equation. It can be shown that the action in Eq.(\ref{GSMaction2}) possesses a hidden general linear group symmetry GL(1,3,R) in coordinate spacetime. By projecting the gauge-type gravitational equation (Eq.(\ref{GaugeGE})) into coordinate spacetime using $\chi_{\mu}^{\; a}$ and $\chih_{a}^{\; \mu}$, we can derive ten symmetric and six antisymmetric gravitational equations:
\be \label{GGE}
& & R_{\mu\nu} -  \frac{1}{2}\chi_{\mu\nu} R  + \gamma_W \cG_{(\mu\nu)} =  8\pi G_{\kappa} \mT_{(\mu\nu)} , \nn \\
& &  \gamma_W \cG_{[\mu\nu]}  = 8\pi G_{\kappa} \mT_{[\mu\nu]} ,
\ee
which are referred to as geometric-type gravitational equations. The symmetric equation provides a generalized Einstein equation due to the presence of tensor $\cG_{\mu\nu}$ and conformal inhomogeneous spin gauge fields, while the antisymmetric equation offers an additional equation, describing gravitational interactions beyond Einstein's GR due to the failure of the equivalence principle. As a consequence, the gravitational waves arising from massless graviton were demonstrated in ref.\cite{GQFTN} to have five independent polarizations instead of two polarizations in GR. The conserved energy momentum tensor is, $\bs{\mT}_{\mu\nu}\equiv -\gamma_W \cG_{(\mu\nu)} + 8\pi G_{\kappa} \mT_{(\mu\nu)}$, with $\nabla^{\mu} \bs{\mT}_{\mu\nu}$ $\equiv \nabla^{\mu} G_{\mu\nu}=0$ due to the Bianchi identity of Einstein's geometric tensor $G_{\mu\nu}$ in GR. The general gravitational equation may be understood from zero energy momentum tensor theorem via the cancellation law of energy momentum tensor in GQFT\cite{GQFT-ZEMT}. The explicit forms for the conserved current, $\whmJ_{a}^{\; \mu} \equiv \gamma_W \hcG_{a}^{\;\mu} + \hsF_{a}^{\;\mu} -16\pi G_{\kappa} \hmJ_{a}^{\;\mu}$, the general tensors $\mT_{\mu\nu}$ and $\cG_{\mu\nu}$, and the relevant derivations are presented in Section C of Supplementary material.

The GL(1,3,R) symmetry, foundational to GR, governs gravity via the dynamics of Riemannian geometry with the gravimetric field $\chi_{\mu\nu}$. In the framework of GQFT, GL(1,3,R) emerges as a hidden symmetry, and gravitational interaction is described by the spin-related gravigauge field $\chi_{\mu}^{\; a}$ with its field strength $\sF_{\mu\nu}^{a}$, associated with the structure factor $\hsF_{cd}^{a}\equiv - \chih_{c}^{\;\mu}\chih_{d}^{\; \nu}\sF_{\mu\nu}^{a}$ in non-commutative gravigauge spacetime. Despite this hidden symmetry and Riemannian geometry, a globally flat Minkowski spacetime can always be chosen as the base spacetime in GQFT. This is because the gravigauge field is identified as massless graviton in the bi-frame spacetime framework. Importantly, gravitational interactions involving spinor fermions (leptons and quarks) and spin gauge bosons occur primarily through the spin-related gravigauge field rather than composite gravimetric field. 

In the above action (Eq.(\ref{GSMaction2})), it is shown explicitly that the gauge fields $\cW_{\mu}^{\; a}$, $\cB_{\mu}$ and $\cW_{\mu}$ do not directly couple to the leptons and quarks due to the chirality and hermiticity properties of the action. A discrete symmetry $Z_2$, $\cW_{\mu}^{\; a}\to -\cW_{\mu}^{\; a}$, $\mW_{\mu}\to - \mW_{\mu}$, with $\mW_{\mu} \equiv \cW_{\mu} + (2g_w)^{-1}\p_{\mu} \ln(\beta_{\kappa}^2\bM_{\kappa}^2 + \phi_w^2 + \phi_e^2 )$, prevents vector gauge bosons $\cW_{\mu}^{\; a}$ and $\mW_{\mu}$ from decaying into other particles, making them stable dark matter candidates. 

In general, all particles receive masses after making appropriate gauge fixing conditions and breaking down spontaneously to the minimum of scalar potentials. However, the scalar potentials cannot be uniquely determined from gauge symmetry. For our present purposes, we assume that all scalar fields can reach their potential minima with non-zero vacuum expectation values (VEVs):  $\phi_w = v_w + \varphi_w$, $\phi_e = v_e + \varphi_e$, and $H^0 = (v_h + h^0)/\sqrt{2}$ for the Higgs boson in the SM. This brings all particles to be massive, except for the gravigauge field, which behaves as Goldstone-type boson. Note that the scaling gauge invariance requires all VEVs arising from dimensionless parameters. For instance, the Higgs boson gets the minimum at $\langle H^2/\Phi_{\kappa}^2 \rangle = \epsilon_h$, with $\epsilon_h$ being the dimensionless parameter. In the fundamental mass scaling gauge basis, $\Phi_{\kappa}=\bM_{\kappa}$, the VEV is given by, $v_h= \sqrt{2\epsilon_h}\bM_{\kappa}$. The smallness of parameter $\epsilon_h$ may be attributed to a tiny number resulting from the ratio of two characteristic energy scales, $\epsilon_{\kappa} \equiv \Lambda_{\kappa}/\bM_{\kappa}$, with $\Lambda_{\kappa}$ the basic cosmological energy scale, $\Lambda_{\kappa}\sim 10^{-3}$ eV. This brings an extremely small number,  $\epsilon_{\kappa} \sim 10^{-30}$. The small VEV $v_h$ is characterized by this tiny ratio via the relation, $\epsilon_h\equiv \epsilon_{\kappa}^{\gamma_h}$, with $\gamma_h \sim 1$. Nevertheless, the naturalness of small mass scale and the stability of scalar potential are the long standing issues, various mechanisms have been explored, such as dynamically spontaneous symmetry breaking\cite{DSSB}, which deserves to be studied further in GSM.

It is useful to analyze the independent degrees of freedom (DoF) for all gauge bosons. The spin gauge symmetry can be fixed by taking a special gauge transformation that transforms the gravigauge field to be symmetric, $\chi_{\mu a} = \chi_{a\mu}$, which allows to remove six redundant degrees of freedom in the spin gauge field and result in massive spin gauge boson. When choosing the unitary conformal-boost gauge basis and fundamental mass scaling gauge basis via transforming vector field and singlet scalar fields into the specific fixing conditions: $\bs{\zeta}^a(x) = 0$, $\bs{\zeta}(x)=1$ and $\Phi_{\kappa}(x) = \bM_{\kappa}$, we are able to eliminate six redundant degrees of freedom, four DoF in the chirality boost-spin gauge field, one DoF in chiral conformal-spin gauge field and one DoF in scaling gauge field, which enables the gauge bosons $\cW_{\mu}^{a}$, $\cB_{\mu}$ and $\mW_{\mu}$ to receive masses.

All unknown parameters in SM have been explored through various experiments. Here, the additional observables include masses of gauge and scalar bosons: $M_{\cA} = g_{4}\sqrt{\gamma_{\kappa}^2\bM_{\kappa}^2 + \beta_e v_e^2 }$, $M_{\mW} = g_{w}\sqrt{\beta_{\kappa}^2\bM_{\kappa}^2 + v_w^2 + v_e^2 }$, $m_{\cB} = g_c\gamma_w v_w$, $m_{\cW} = g_4 \beta_w v_w$, $m_{\varphi_w} = \lambda_{w} v_w$ and $ m_{\varphi_e} = \lambda_{e} v_e$, with $v_w$ and $v_e$ two VEVs, and $\lambda_w$ and $\lambda_e$ two constants in potentials, which correspond to the masses of the boson fields $\cA_{\mu}^{ab}$, $\mW_{\mu}$, $\cB_{\mu}$, $\cW_{\mu}^{a}$, $\varphi_w$ and $\varphi_e$.

Let us discuss some interesting properties of gauge boson $\cW_{\mu}^{a}$ as dark matter candidate. In general, its mass can only be determined by experiment as it depends on two free parameters $v_w$ and $\beta_w$. While the mass region may fall into the present and future experimental sensitivities, ranging from TeV to MeV, if the VEV $v_w$ is characterized, analogous to Higgs VEV $v_h$ in SM, by the small number, $\epsilon_{\kappa}\equiv \Lambda_{\kappa}/\bM_{\kappa}\sim 10^{-30}$, via the relation, $v_w= \sqrt{2\epsilon_w}\bM_{\kappa}$, $\epsilon_w\equiv \epsilon_{\kappa}^{\gamma_w}$, with $\gamma_w = 1.0\sim 1.2$. The gauge boson $\cW_{\mu}^{a}$ interacts with all leptons and quarks through heavy spin gauge boson $\cA_{\mu}^{ab}$, its interaction strength is governed by the spin gauge boson mass $M_{\cA}$, which depends on free parameters $g_4$ and $\gamma_{\kappa}$, multiplied by the fundamental mass $\bM_{\kappa}$ at Planck scale. Unlike the usual dark matter models, the gauge boson $\cW_{\mu}^{a}$ behaves as bi-covariant vector field, analogous to the massless graviton $\chi_{\mu}^{\; a}$, may be referred to as dark graviton for convenience. A detailed study on the Weyl scaling gauge boson as dark matter candidate was presented in ref. \cite{TW}. Its mass depends on the free parameters $g_w$ and $\beta_{\kappa}$, multiplied by the fundamental mass $\bM_{\kappa}$. 

The gauge boson $\cB_{\mu}$ and the scalar boson $\varphi_w$ could also serve as dark matter candidates, depending on their masses and decay rates. If their masses are less than those of dark graviton and lightest neutrino, determined by the parameters $\gamma_w$ and $\lambda_w$, or if their decay rates via dark graviton and spin gauge boson are extremely low, resulting in lifetimes much longer than the age of the universe, they can become viable dark matter candidates.      

The scalar field $\phi_e$ is proposed to play a dual role as sources of both primordial energy and dark energy. To achieve this, let us express its potential as: $\cV_e(\phi_e)\equiv \lambda_E^4 \bM_{\kappa}^4\cF_e(\chi_e)$, where $\chi_e\equiv \phi_e/(\lambda_{\kappa}\bM_{\kappa})$. The function $\cF_e(\chi_e)$ must satisfy slow-roll and finiteness conditions, specifically, $\cF_e(\chi_e) \to \lambda_P^4$ for $\phi_e\gg\lambda_{\kappa}\bM_{\kappa}$. The potential reaches its minimum when $\cF_e'(\chi_e) =0 $ at $\phi_e=v_e$, with $\cF_e(v_e) \simeq \Lambda_{\kappa}^4/\bM_{\kappa}^4 \ll 1$ and $v_e\simeq \lambda_{\kappa}\Lambda_{\kappa}^2/\bM_{\kappa}$. Here, $\lambda_E$, $\lambda_P$ and $\lambda_{\kappa}$ are free parameters. This allows the scalar boson to act as the source of primordial energy, $\Lambda_{PE}^4 \simeq \lambda_E^4 \lambda_P^4 \bM_{\kappa}^4$, driving early inflation, and also as the source of dark energy, $\Lambda_{DE}^4 \simeq \lambda_E^4 \Lambda_{\kappa}^4$, driving the ongoing accelerated expansion. Meanwhile, the canonical scalar boson $\varphi_e$ acquires a tiny cosmic mass, $m_{\varphi_e} \propto \lambda_E^2 \lambda_P \lambda_{\kappa}^{-1}v_e$, which may be referred to as dark cosmino. A typical potential may simply be taken as $\cF_e(\chi_e)= (\chi_e^4 + \Lambda_{\kappa}^4/\bM_{\kappa}^4)/(1 +\lambda_P^{-2} \chi_e^2)^2$. Unlike the simplest dark energy model with a pure cosmological constant, such a dark cosmino enables to describe the dynamic feature of dark energy and bring a richer structure as exhibited in a teleparallel dark energy model\cite{TDE}.

In conclusion, the general theory of standard model (GSM), grounded on conformal inhomogeneous spin gauge and scaling gauge symmetries WS$_c$(1,3)$\times$SG(1) along with SM symmetry, has been built to provide a unified description of fundamental forces and comprehensive understanding on the nature of spacetime.  The GSM surpasses GR and SM in describing gravidynamics and addresses the mysteries of the dark side of the universe.

\centerline{{\bf Acknowledgement}}

This work was supported in part by the National Science Foundation of China(NSFC) under Grants No.~12347103 and No.~11821505, the National Key Research and Development Program of China under Grant No.2020YFC2201501, and Strategic Priority Research and special fund of the Chinese Academy of Sciences.


\vspace{1.0cm}

{\bf{Appendix A. Supplementary materials}}

\subsection{A General  Theory of Standard Model (GSM)}

The action for the general theory of standard model is explicitly constructed  in a spin-related intrinsic gravigauge spacetime as follows: 
\be \label{GSMaction0}
& & \cS_{GSM}  = \int [\dbar \zeta^{c}]  \sum_{s=\mp}  \frac{1}{2} \{  \sum_{\Psi=l,q}  \bs{\zeta} [ \frac{1}{2} \bar{\Psi}_{s}^{i} \Sigma_{s}^{c}  i \bshcD_{c}^{(\Psi s)}\, \Psi_{s}^{i}  + H.c. 
 \nn \\
& & + \bar{\Psi}_{s}^{i} \hPhi_{s} ( \tilde{\Gamma}_{s} \tilde{\lambda}^{\Psi}_{ij}  +  \hGa_{s}  \hat{\lambda}^{\Psi}_{ij} ) \, \Psi_{s}^{j} ] + \frac{1}{8} \eta^{cd} \Tr (\hcD_{c}\hPhi_{s} )^{\dagger} \hcD_{d}\hPhi_{s} \nn \\
& & -  \frac{1}{4} \eta^{cc'}\eta^{dd'} ( \frac{1}{2g_1^2} \Tr \hF_{cd}^{(l_{s})} \hF_{c'd'}^{(l_{s})}  + \frac{1}{2g_2^2} \Tr \hF_{cd}^{(L_{s})} \hF_{c'd'}^{(L_{s})} )   \} \nn \\
& & -\frac{1}{4}\eta^{cc'}\eta^{dd'} [ \frac{4}{g_3^2} \Tr \hF_{cd}^{(C)} \hF_{c'd'}^{(C)} + \frac{1}{g_w^2} \hcW_{cd}\hcW_{c'd'}  \nn \\
& & +\frac{1}{2g_4^2} \Tr (\bshcF_{cd}^{-} \bshcF_{c'd'}^{-} + \bshcF_{cd}^{+} \bshcF_{c'd'}^{+}   \nn \\ 
& & -  \frac{1}{2\bs{\zeta}^{2}} \bshcF_{cd}^{(W-)} \bshcF_{c'd'}^{(W+)} + \frac{2}{\gamma_c^2} \hcF_{cd}^{(-)} \hcF_{c'd'}^{(+)}  ) ]\nn \\
& & +  \frac{1}{4} \Phi_{\kappa}^2  \etat^{cdc'd'}_{aa'} \bshsF_{cd}^{a}\bshsF_{c'd'}^{a'}    - \frac{1}{8} \beta_w^2 \hphi_w^2   \eta^{cd}  \Tr \bshcW_{c}^{(-)}  \bshcW_{d}^{(+)}  \nn \\
& & +   \frac{1}{2}  (  \gamma_{\kappa}^2 \Phi_{\kappa}^2  + \beta_e \hphi_{e}^2 )  \eta^{cd}  \Tr \bshcA_{c}\bshcA_{d} + \frac{1}{2}\beta_{\kappa}^2 \eta^{cd}\cD_{c}\Phi_{\kappa} \cD_{d}\Phi_{\kappa}  \nn \\
& & +\frac{1}{2}  \eta^{cd} [ \hcD_{c}(\bs{\zeta}\hphi_{w})\hcD_{d}(\bs{\zeta}\hphi_{w})   +  \gamma_{w}^2  \hphi_w^2\hcD_{c}\bs{\zeta} \hcD_{d}\bs{\zeta} + \cD_{c}\hphi_{e} \cD_{d}\hphi_{e}] \nn \\
& & - \hPhi_{\kappa}^4\, \hcV_{S}( \hPhi/\hPhi_{\kappa}, \bs{\zeta} \hphi_w/\hPhi_{\kappa}, \hphi_{e}/\hPhi_{\kappa}) ,
\ee
with Latin indices ($a,b,c,d=0,1,2,3$) representing spin-related vectors, raised and lowered, and contracted by metric matrices, $\eta^{ab}$($\eta_{ab}$)=$\diag.$(1,-1,-1,-1). Here, the summations over $s=\mp$ and $\Psi=l,q$ are in respective to the chiral duality representations and spinor fields of leptons and quarks. We have defined gauge covariant derivatives,
\be
& & i\bshcD_{c}^{(\Psi \mp)} \equiv i \eth_{c}+ \hA_{c}^{(\Psi \mp)} + \bshcA_{c}^{\mp} + i 3\hcW_{c}/2, \nn \\
& & \hA_{c}^{(\Psi \mp)}\equiv  \hB_{c} \Sigma_{Y \mp}^{(\Psi)} + \hW_{c}^{i} \Sigma_{L \mp}^i  + \hA_{c}^{\alpha} T_{\alpha}^{(\Psi)}, \nn \\
& & \bshcA_{c}^{\mp} \equiv \hcB_{c}\Sigma_{\mp} + \hcW_{c}^{a}\Sigma_{a \mp} + \hcA_{c}^{ab}\Sigma_{ab}/2 , 
\ee
where $\hB_{c}$, $\hW_{c}^{i}$, and $\hA_{c}^{\alpha}$ correspond to the electroweak and strong gauge fields in SM, $\hcB_{c}$, $\hcW_{c}^{a}$ and $\hcA_{c}^{ab}$ are chiral conformal-spin, chirality boost-spin and spin gauge fields introduced via gauge symmetry WS$_c$(1,3), and $\hcW_{c}$ is Weyl scaling gauge field. We have defined the gauge covariant fields, 
\be \label{GCGF}
& & \bshcW_{c}^{(\mp)}  \equiv (\hcW_{c}^{\; a}  - \hcD_{c}\bs{\zeta}^{a} )  \Sigma_{\mp a}, \nn \\
& & \bshcA_{c} \equiv (\hcA_{c}^{ab}  -  \bshmOm_{c}^{ab} )\Sigma_{ab}/2,
\ee
with the covariant derivative over $\bs{\zeta}^{a}$, $\hcD_{c}\bs{\zeta}^{a} = (\eth_{c} +  \hcB_{c}) \bs{\zeta}^{a}  + \hcA_{c b}^{a}\bs{\zeta}^{b}$. Here, a spin-related vector field $\bs{\zeta}^{a}$ is introduced to ensure the gauge covariance. 

The scalar fields $\hPhi_{\mp} = \hphi_{I} \Sigma_{h \mp}^{I}$ with $\Sigma_{h -}^{I}\equiv \Gamma^{I}\Gamma_{-}$ and $\Sigma_{h +}^{I}\equiv -(\Gamma^{I})^{\ast}\Gamma_{+}$ represent Higgs boson, and four scalars $\hphi_{I}$ ($I=5,6,7,8$) constitute Higgs doublet $\hH$ in SM with the charged and neutral Higgs bosons, 
\be
\hH^{+} \equiv \hphi_7 + i \hphi_6, \quad \hH^{0} \equiv \hphi_5 - i \hphi_8 .
\ee

The $\Sigma$-matrices $\Sigma_{\mp}^{A}\equiv\Gamma^{A} \Gamma_{\mp}$ are defined via $\Gamma$-matrices $\Gamma^{A}\equiv(\Gamma^a , \Gamma^{I})$($a=0,1,2,3$, $I=5,6,7,8$) with the chiral project $\Gamma$-matrices $\Gamma_{\mp}$. The $\Gamma$-matrices $\Gamma^{A}\equiv(\Gamma^a , \Gamma^{I})$($a=0,1,2,3$, $I=5,6,7,8$) and relevant chirality $\Gamma$-matrices in the action (Eq.(\ref{GSMaction0})) are defined as follows: 
\be
& & \Gamma^{a} = \, \sigma_0 \otimes \sigma_0 \otimes \gamma^{a} ,  \; \;  \Gamma^5 =  i\sigma_2\otimes \sigma_0\otimes \gamma_5 , \nn\\
& &  \Gamma^6 = i\sigma_1\otimes \sigma_1 \otimes\gamma_5, \;\;  \gamma_9 = \sigma_3\otimes \sigma_0 \otimes\gamma_5 , \nn \\
& & \Gamma^7 = i\sigma_1\otimes \sigma_2 \otimes\gamma_5,\;\; \tilde{\gamma}_9 =  \sigma_3\otimes \sigma_0 \otimes I_4, \nn \\
& & \Gamma^8 = i\sigma_1\otimes \sigma_3 \otimes\gamma_5, \;\;  \hat{\gamma}_9 = \sigma_0\otimes \sigma_3 \otimes I_4,  \nn \\
& & \Gamma_{\mp} = ( 1 \mp\gamma_9)/2, \;\; \tilde{\Gamma}_{\mp} = ( 1 \mp \tilde{\gamma}_9)/2 ,
\ee
with $\gamma^{a} $ and $\gamma_5=\sigma_3\otimes \sigma_0$ as Dirac-type $\gamma$-matrices.
The chiral project matrices, $\tilde{\Gamma}_{\mp} $ and $\hat{\Gamma}_{\mp} =  \hat{\gamma}_9 \tilde{\Gamma}_{\mp}$ are adopted to obtain the Higgs doublet couplings in SM.  

The group generators of SM symmetry have the following explicit forms in chiral spinor representations:  
\be \label{GG1}
& & \Sigma_{Y \mp}^{(\Psi)} \equiv (N^{(\Psi)} +  \hat{\Gamma}_{\mp})/2 , \; N^{(l)}= -1,\;  N^{(q)}=1/3, \nn \\
& & \Sigma_{L\mp}^i \equiv \Sigma^i  \tilde{\Gamma}_{\pm} , \; T_{\alpha}^{(l)}=0, \; T_{\alpha}^{(q)}=\lambda_{\alpha}/2 ,\nn \\
& & \Sigma^i\equiv \Sigma^{I-5} = \epsilon^{IJK}\Sigma^{JK},  \;\; \Sigma^{JK} = i[ \Gamma^J, \Gamma^K ]/4, 
\ee
with $\Sigma_{Y}^{(\Psi)}$, $\Sigma_{L\mp}^{i}$ ($i=1,2,3$, $I,J,K=6,7,8$) and $T_{\alpha}^{(q)}$ ($\alpha=1,\cdots,8$) corresponding to the group generators of U$_Y$(1), SU$_L$(2) and SU$_c$(3). 

The group generators for the conformal inhomogeneous spin gauge symmetry WS$_c$(1,3) in GSM are defined explicitly by the $\Gamma$-matrices as follows:
\be
& & \Sigma^{a b} = i [\Gamma^{a}, \Gamma^{b} ]/4 ,  \; \Sigma_{\mp}^{a}= \Gamma^{a} \Gamma_{\mp} , \; \Sigma_{\mp} = \pm i\gamma_9/2 , 
\ee
 where the generators ($\Sigma^{ab}, \Sigma_{-}^{a}, \Sigma_{-})$$\equiv \Sigma_{-}^{\ta \tb} $  and the chiral dual ones, $(\Sigma^{ab}, \Sigma_{+}^{a}, \Sigma_{+})$$\equiv\Sigma_{+}^{\ta \tb}$, with $\Sigma_{+}^{\ta \tb} = C_{d}\Sigma_{-}^{\ta \tb}C_{d}^{-1}$, correspond to the group generators of SP(1,3), $W^{1,3}$ and SP(1,1) gauge symmetries, they satisfy the following group algebra of WS$_c$(1,3):
\be
& & [\Sigma^{ab}, \Sigma^{cd}] =  i (\Sigma^{ad}\eta^{bc} -\Sigma^{bd}  \eta^{ac} - \Sigma^{ac} \eta^{bd} + \Sigma^{bc} \eta^{ad}) , \nn \\
& &  [\Sigma^{ab}, \Sigma_{\mp}^{c}] = i( \Sigma_{\mp}^{a}\eta^{bc} -\Sigma_{\mp}^{b}  \eta^{ac}  ) , \quad [\Sigma_{\mp}, \Sigma_{\mp}^{a}] = i \Sigma_{\mp}^{a} , \nn \\
& &  [\Sigma^{ab}, \Sigma_{\mp} ]  = 0,  \quad [\Sigma_{\mp}, \Sigma_{\mp} ]=0, \quad [\Sigma_{\mp}^{a}, \Sigma_{\mp}^{b}] = 0 .
\ee

The commutation relation of covariant derivative $i\bshD_{c} = i\eth_{c} + \bshA_{c}$ is found to be,
\be
[ i\bshD_c ,\; i\bshD_d] = i ( \bshF_{cd} + \hsF_{cd}^a\, i\bshD_a ) ,
\ee
with the gauge covariant field strength $\bshcF_{cd}$ given by,
\be
\bshF_{cd} & \equiv & \bsF_{cd} - \hsF_{cd}^{a} \bshA_{a} , \nn \\
 \bsF_{cd} & \equiv & \eth_{c}\bshA_{d} - \eth_{d}\bshA_{c} - i [\bshA_{c}, \bshA_{d}],  
\ee  
where the term $\hsF_{cd}^{a}\bshA_{a}$ reflects gravitational effect. To obtain scaling gauge covariant field strength $\bshF_{cd}$, it is useful to introduce the scaling gauge covariant field strength $\bshsF_{cd}^{a}$ and gravigauge derivative $\hat{\eth}_c$ as follows:
\be
\bshsF_{cd}^{a} & \equiv & \hsF_{cd}^{a} + \hcS_{[cd]}^{a} , \;\; \bshmOm_{c}^{ab}  \equiv \hmOm_{c}^{ab}   - \hcS_{c}^{[ab]} , \nn \\
\hat{\eth}_c & \equiv & \eth_c + \cS_c , \quad \cS_c \equiv \eth_{c}\ln \hPhi_{\kappa}, 
\ee
where we have the following relations and definitions:
\be
\hsF_{cd}^{a} & \equiv & \hmOm_{cd}^{a} - \hmOm_{dc}^{a},  \quad \hmOm_{c}^{ab} = \etach_{ca'}^{[ab] c'd'} \hsF_{c'd'}^{a'} , \nn \\ 
\hcS_{[cd]}^{a} & \equiv & \eta_{[cd]}^{[ab]}\cS_b, \; \hcS_{c}^{[ab]} \equiv \eta_{[cd]}^{[ab]}\cS^d,
\ee
with 
\be
\etach_{ca'}^{[ab] c'd'} & \equiv&  \frac{1}{2}(\eta^{ac'}\eta_{a'}^{\, b} - \eta^{bc'}\eta_{a'}^{\, a}) \eta_{c}^{\, d'} \nn \\
& + & \frac{1}{4}(\eta^{ac'}\eta^{bd'} - \eta^{bc'}\eta^{ad'} ) \eta_{ca'}, \nn \\
\eta_{[cd]}^{[ab]} & \equiv &  \eta_{c}^{\, a} \eta_{d}^{\, b}  - \eta_{d}^{\, a} \eta_{c}^{\, b} .
\ee
Here, $\hmOm_{c}^{ab}$ is referred to as spin gravigauge field, related to $\hsF_{cd}^{a}$ through the tensor $\etach_{ca'}^{[ab] c'd'}$. $\cS_c$ is regarded as a pure gauge field via the basic scalar field $\hPhi_{\kappa}$. 

 A special tensor $\tilde{\eta}^{cd c'd'}_{a a'}$ in the action Eq.(\ref{GSMaction0}) is constructed to ensure the spin gauge invariance,
 \be
 \tilde{\eta}^{cd c'd'}_{a a'}  &\equiv& \eta^{c c'} \eta^{d d'} \eta_{a a'}  +  \eta^{c c'} ( \eta_{a'}^{d} \eta_{a}^{d'}  -  2\eta_{a}^{d} \eta_{a'}^{d'}  ) \nn \\
 & + & \eta^{d d'} ( \eta_{a'}^{c} \eta_{a}^{c'} -2 \eta_{a}^{c} \eta_{a'}^{c'} ).
 \ee

All relevant gauge covariant field strengths, $\bshF_{cd}\equiv(\hF_{cd}^{(l_{\mp})}, \hF_{cd}^{(L_{\mp})}, \hF_{cd}^{(C)}, \hcW_{cd},  \hcF_{cd} , \hcF_{cd}^{(\mp)}, \hcF_{cd}^{W\mp})\equiv(\hF_{cd}\Sigma_{Y \mp}^{(l)} , \hF_{cd}^{i} \Sigma_{L\mp}^i, \hF_{cd}^{\alpha}T^{\alpha} , \hcW_{cd}, \hcF_{cd}^{ab}\Sigma^{ab}/2, \hcF_{cd} \Sigma_{\mp},\hcF_{cd}^{a}\Sigma_{\mp a})$, and $\bshcF_{cd}^{(W\mp)}$ in the action Eq.(\ref{GSMaction0}) are defined as follows:
\be
& & \hF_{cd}^{(\Psi_{\mp})} \equiv \hF_{cd}^{(l_{\mp})}  + \hF_{cd}^{(L_{\mp})} + \hF_{cd}^{(C)} , \nn \\
& & \hF_{cd}^{(l_{\mp})} \equiv (\hat{\eth}_{c} \hB_{d} - \hat{\eth}_{d}\hB_{c} + \bshsF_{cd}^{a}\hB_{a}  ) \Sigma_{Y \mp}^{(l)}, \nn \\
& & \hF_{cd}^{(L_{\mp})} \equiv  (\hat{\eth}_{c} \hW_{d}^{i} - \hat{\eth}_{d}\hW_{c}^{i} + \epsilon^{ijk}\hW_{c}^{j}  \hW_{d}^{k} + \bshsF_{cd}^{a} \hW_{a}^{i} )  \Sigma_{L\mp}^i , \nn \\
& & \hF_{cd}^{(C)} \equiv (\hat{\eth}_{c} \hA_{d}^{\alpha} - \hat{\eth}_{d}\hA_{c}^{\alpha} -i  [\hA_{c}, \hA_{d}]^{\alpha} + \bshsF_{cd}^{a}\hA_{a}^{\alpha} ) \lambda^{\alpha}/2 , \nn \\
& & \hcW_{cd}\equiv  \hat{\eth}_{c} \hcW_{d} - \hat{\eth}_{d}\hcW_{c}  + \bshsF_{cd}^{a} \hcW_{a}  ,
\ee
and
\be  
& &  \bshcF_{cd}^{\mp} \equiv  \hcF_{cd} +  \bshcF_{cd}^{(W\mp)} + \hcF_{cd}^{(\mp)}, \nn \\
& &  \hcF_{cd} \equiv (\hat{\eth}_{c} \hcA_{d}^{ab} - \hat{\eth}_{d}\hcA_{c}^{ab} -i [\hA_{c}, \hA_{d}]^{ab} + \bshsF_{cd}^{a'} \hcA_{a'}^{ab} ) \Sigma^{ab}/2 , \nn \\
& & \hcF_{cd}^{W\mp} \equiv ( \hcD_{c} \hcW_{d}^{a} - \hcD_{d} \hcW_{c}^{a}) \Sigma_{\mp a} , \nn \\
& &  \hcF_{cd}^{(\mp)} \equiv  (\hat{\eth}_{c} \hcB_{d} - \hat{\eth}_{d}\hcB_{c} + \bshsF_{cd}^{a} \hcB_{a} ) \Sigma_{\mp}, \nn \\
& & \bshcF_{cd}^{(W\mp)} \equiv  (\hcF_{cd}^{a} - \hcF_{cd}\bs{\zeta}^{a}  - \hcF_{cd}^{ab} \bs{\zeta}_{b} ) \Sigma_{\mp a} ,
\ee
where $\bshcF_{cd}^{(W\mp)}$ is the field strength of $\bshcW_{c}^{(\mp)}$ defined in Eq.(\ref{GCGF}). The covariant derivative $\hcD_{c} \hcW_{d}^{a}$ in the definition of the field strength $\hcF_{cd}^{W\mp} \equiv \hcF_{cd}^{a}\Sigma_{\mp a}$ is given by,
\be
& & \hcD_{c} \hcW_{d}^{a}= (\hat{\eth}_{c} + \hcB_{c})\hcW_{d}^{a}  +  \bshmOm_{c d}^{b}\hcW_{b}^{a} + \hcA_{c b}^{a}\hcW_{d}^{b} .
\ee
The covariant derivative over $\hPhi_{\mp}$ is defined as follows:
\be
& & \hcD_{c}\hPhi_{\mp} \equiv (\hat{\eth}_{c} - i \hB_{c} \tGa_{\pm}/2  - i \hW_{c}^{i}\Sigma_{L \mp}^{i} ) \hPhi_{\mp}  ,  \nn \\
& &  \eta^{cd} \Tr (\hcD_{c}\hPhi_{\mp} )^{\dagger} \hcD_{d}\hPhi_{\mp} = 8\eta^{cd} (\hcD_{c}\hH)^{\dagger} \hcD_{d}\hH .
\ee

\subsection{ Gravitization Equation of GSM in spin-related Gravigauge Spacetime}

The matrices $\bscM_{cda}^{\; c'd'a'}$ and $\widehat{\bscF}_{cda}$ in the gravitization equation are explicitly  defined as follows: 
\be
& & \bscM_{cda}^{\; c'd'a'}  \equiv (\eta_{c}^{\, c'} \eta_{d}^{\, d'} -\eta_{d}^{\, c'} \eta_{c}^{\, d'} )/2 ( \eta_{a}^{\, a'} \hat{\cS}_{+}^2   - \widehat{\cV}_{a b} \eta^{b a'} ) \nn \\
& & +  [ ( \eta_{c}^{\, c'}  \eta_{d}^{\, a'}  - \eta_{d}^{\, c'} \eta_{c}^{\, a'} ) \eta_{a}^{\, d'} +  (\eta_{d}^{\, d'} \eta_{c}^{\, a'} - \eta_{c}^{\, d'} \eta_{d}^{\, a'} )\eta_{a}^{\, c'} ] \hat{\cS}_{-}^{2}/2 \nn \\
& & -  [ \eta_{ca} (\eta_{d}^{\, d'}  \eta^{c'a'} - \eta_{d}^{\, c'}  \eta^{d'a'}) -  \eta_{da}( \eta_{c}^{\, d'}  \eta^{c'a'} -\eta_{c}^{\, c'}  \eta^{d'a'} ) ] \Phi^2_{\kappa} , \nn \\
& & \widehat{\cV}_{a b}  \equiv   \hcB_{a} \hcB_{b} - \hcW_{a c} \hcW_{b}^{c} + \hcA_{a a'b'}\hcA_{b}^{a'b'}   + \hcW_{a} \hcW_{b} \nn \\
& & \qquad + \hB_{a}\hB_{b}  + \hW_{a}^{i} \hW_{b}^{i} + \hA_{a}^{\alpha} \hA_{b}^{\alpha } = \widehat{\cV}_{ba} , 
\ee
with $\hat{\cS}_{\pm}^2  =   \Phi_{\kappa}^2 + \frac{1}{2} (\gamma_{\kappa}^2\Phi^2_{\kappa} + \beta_{e}\hphi_{e}^2)  \pm  (\gamma_{\kappa}^2\Phi^2_{\kappa} + \beta_{e}\hphi_{e}^2)$, and 
\be
\widehat{\bscF}_{cda}  & \equiv &   \cF_{cd}\hcB_{a} -  \cF_{cd}^{b} \hcW_{a b}  + \cF_{cd}^{a'b'}\hcA_{a a'b'}    \nn \\
& + &  \cW_{cd} \hcW_{a} + F_{cd} \hB_{a} +  F_{cd}^{i} \hW_{a}^{i} + F_{cd}^{\alpha}\hA_{a}^{\alpha} \nn \\
& - &  (\gamma_{\kappa}^2\Phi^2_{\kappa} + \beta_{e}\hphi_{e}^2) g_4( 2\hcA_{cda}-2\hcA_{dca} +\hcA_{acd} )  \nn \\
& + & [ 4\Phi_{\kappa}^2 - 3(\gamma_{\kappa}^2\Phi^2_{\kappa} + \beta_{e}\hphi_{e}^2) ] \hcS_{[cd]a}  =  - \widehat{\bscF}_{dc a}  ,
\ee
where $\bscM_{cda}^{\; c'd'a'}$ is regarded as a $24\times 24$ matrix, antisymmetric under exchanges of $c, d$ and $c', d'$, and $\widehat{\bscM}_{cda}^{\; c'd'a'}$ is the inverse matrix. Note that the gauge fields are rescaled as, $\hcA_{c}^{A} \to g_A \hcA_{c}^{A}$, in order to normalize their kinetic terms.

\subsection{Gravitational Equations of GSM within Gravitational Quantum Field Theory(GQFT)}

In obtaining the action within the framework of GQFT, we have used the following relation,
\be \label{Relation}
& &  \eta^{cd} ( \cA_{c}^{ab}-\mOm_{c}^{ab}/g_4 )  ( \cA_{d ab}-\mOm_{d ab}/g_4) \equiv  \frac{1}{2 g_4^2} \bchi^{\mu\nu \mu' \nu'}_{aa'}\cG_{\mu\nu}^{a}\cG_{\mu'\nu'}^{a'}, \nn \\
& & \cG_{\mu\nu}^{a} \equiv \p_{\mu} \chi_{\nu}^{\; a}  - \p_{\nu} \chi_{\mu}^{\; a} + g_4(\cA_{\mu b}^{a} \chi_{\nu}^{\; b}  - \cA_{\nu b}^{a} \chi_{\mu}^{\; b} ),  
\ee
with the definitions,
\be \label{Tensor}
& & \etab^{c d c' d'}_{a a'} \equiv  \frac{3}{2} \eta^{c c'} \eta^{d d'} \eta_{a a'}  - \frac{1}{2} ( \eta^{c c'} \eta_{a'}^{d} \eta_{a}^{d'}  +  \eta^{d d'} \eta_{a'}^{c} \eta_{a}^{c'} ), \nn \\
& & \chib_{aa'}^{\mu\nu \mu'\nu'} \equiv \chih_{c}^{\;\, \mu}\chih_{d}^{\;\, \nu} \chih_{c'}^{\;\, \mu'} \chih_{d'}^{\;\, \nu'}  \etab^{c d c' d'}_{a a'}.
\ee

The current $\whmJ_{a}^{\;\mu}$ in the gauge-type gravitational equation can be expressed into the following explicit forms: 
\be \label{CC}
& & \whmJ_{a}^{\;\mu}\equiv \gamma_W \hcG_{a}^{\;\mu}  + \hsF_{a}^{\;\mu} -16\pi G_{\kappa} \hmJ_{a}^{\;\mu} , \nn \\
& & \hcG_{a}^{\;\mu} = - \cD_{\nu}(\chi_{\kappa}^2 \widehat{\cG}_{a}^{\mu\nu} ) + \chi_{\kappa}^2  (\eta_{\sigma}^{\; \mu}\chih_{a}^{\; \rho} -  \frac{1}{4} \chih_{a}^{\; \mu} \eta_{\sigma}^{\; \rho} ) \cG_{\rho\nu}^{b} \widehat{\cG}^{\sigma\nu}_{b} ] , \nn \\
& & \hsF_{a}^{\;\mu} =  (\eta_{\sigma}^{\; \mu}\chih_{a}^{\; \rho} -  \frac{1}{4} \chih_{a}^{\; \mu} \eta_{\sigma}^{\; \rho} ) \sF_{\rho\nu}^{b} \whsF^{\sigma\nu}_{b},
\ee
and
\be \label{CC1}
& & \hmJ_{a}^{\;\mu} =  \chi \1 \{  \sum_{\Psi=l,q} \frac{1}{2} [  (   \chih_{a}^{\; \; \mu} \chih_{b}^{\; \nu} -\chih_{b}^{\; \mu}  \chih_{a}^{\;\nu} )  \bar{\Psi}_{L}^{i} \gamma^{b} i\cD_{\nu}^{(\Psi_L)} \Psi_{L}^{i}  \nn \\
& &\, + \chih_{a}^{\;\;\mu}( \bar{\Psi}_{L}^{i} H \lambda^{\Psi^{(-)}}_{ij} \Psi_{R}^{(-)j} +  \bar{\Psi}_{L}^{i} \tilde{H} \lambda^{\Psi^{(+)}}_{ij} \Psi_{R}^{(+)j} + H.c. ) ] \nn \\
& & \, + ( \chih^{\mu\sigma} \chih_{a}^{\; \; \rho} - \frac{1}{4} \chih_{a}^{\; \; \mu}  \chih^{\rho\sigma} ) ( \sum_{\cF^A} \bscF_{\rho\nu}^{A} \bscF_{\sigma\nu'}^{A} - \cF_{\rho\nu}^{b} \cF_{\sigma\nu' b}) \chih^{\nu\nu'} \nn \\
& &\,  - ( \chih^{\mu\sigma} \chih_{a}^{\; \; \rho} - \frac{1}{2} \chih_{a}^{\; \; \mu}  \chih^{\rho\sigma} ) [\cD_{\rho}\phi_{w} \cD_{\sigma}\phi_{w} + \cD_{\rho}\phi_{e} \cD_{\sigma}\phi_{e} \nn \\
& &\,   + \beta_{\kappa}^2 g_w^2 \bM_{\kappa}^2 \cW_{\rho}\cW_{\sigma} + g_4^2  \phi_w^2 ( \gamma_w^2 \cB_{\rho} \cB_{\sigma} - \beta_{w}^2 \cW_{\rho}^{a}  \cW_{\sigma a}) \nn \\
& & \, + 2(\cD_{\rho}H)^{\dagger} \cD_{\sigma}H ] -  \chih_{a}^{\; \; \mu} \cV_{S}(H, \phi_w, \phi_{e}) \},
\ee
where we have used the following definitions,
\be
 \whcG_a^{\mu\nu} & \equiv & \chi\, \bchi^{[\mu\nu]\mu'\nu'}_{a a'}  \cG_{\mu'\nu' }^{a'}, \nn \\
 \chi_{\kappa}^{2} & \equiv & 1 + g_4^2\beta_e (\phi_e^2-v_e^2)/M_{\cA}^2 , \nn \\
 \gamma_W & \equiv & M_{\cA}^2/(g_4^2 \bM_{\kappa}^2), \quad 16\pi G_{\kappa} = 1/\bM_{\kappa}^2 ,
\ee
with $M_{\cA}$ being the mass of spin gauge boson. 

In the action of GSM, all covariant field strengths $\bscF_{\mu\nu}^{A}$ in Minkowski spacetime are related to those $\bscF_{cd}^{A}$ in spin-related gravigauge spacetime via the projections, ($F_{\mu\nu}$, $ F_{\mu\nu}^{i}$, $F_{\mu\nu}^{\alpha}$, $\cW_{\mu\nu}$, $\cF_{\mu\nu}$, $\cF_{\mu\nu}^{a}$,$\cF_{\mu\nu}^{ab}$) $\equiv$ $\chi_{\mu}^{\; c}\chi_{\nu}^{\; d}$($F_{cd}$, $F_{cd}^{i}$, $F_{cd}^{\alpha}$, $\cW_{cd}$, $\cF_{cd}$, $\cF_{cd}^{a}$, $\cF_{cd}^{ab}$).
Here the covariant field strengths in Minkowski spacetime become the standard ones, 
\be
& & F_{\mu\nu} = \p_{\mu} B_{\nu} - \p_{\nu}B_{\mu} , \nn \\
& & F_{\mu\nu}^i = \p_{\mu} W_{\nu}^i - \p_{\nu}W_{\mu}^i + g\epsilon^{ijk}W_{\mu}^{j}  W_{\nu}^{k}, \nn \\
& & F_{\mu\nu}^{\alpha} = \p_{\mu} A_{\nu}^{\alpha}- \p_{\nu}A_{\mu}^{\alpha} 
- g_3 i [A_{\mu}, A_{\nu}]^{\alpha} , 
\ee
for the gauge fields in SM, and
\be
& &  \cW_{\mu\nu} = \p_{\mu} \cW_{\nu} - \p_{\nu}\cW_{\mu},\nn \\
& & \cF_{\mu\nu} = \p_{\mu} \cB_{\nu} - \p_{\nu}\cB_{\mu},  \nn \\
& & \cF_{\mu\nu}^{a} = (\p_{\mu} + g_c\cB_{\mu}) \cW_{\nu}^{\; a}  - (\p_{\nu} + g_cB_{\nu}) \cW_{\mu}^{\; a} \nn \\
& & \quad \quad + g_4(\cA_{\mu b}^{a} \cW_{\nu}^{\; b}  - \cA_{\nu b}^{a} \cW_{\mu}^{\; b} ),  \nn \\
& &  \cF_{\mu}^{ab} =\p_{\mu} \cA_{\nu}^{ab} - \p_{\nu}\cA_{\mu}^{ab}+ g_4 (\cA_{\mu c}^{a} \cA_{\nu}^{cb} - \cA_{\mu c}^{b} \cA_{\nu}^{ca}) ,
\ee
for the additional gauge fields in GSM.

To understand the geometric-type gravitational equation, it is useful to check that the action term for the gravigauge field strength $\sF_{\mu\nu}^{a}$ is equivalent to the Einstein-Hilbert action up to a total derivative, 
\be
\frac{1}{4} \chi\, \tchi_{aa'}^{\mu\nu \mu'\nu'} \sF_{\mu\nu}^{a} \sF_{\mu'\nu'}^{a'} = \chi\, R
- 2 \p_{\mu} (\chi \chih^{\mu\rho} \chih_{a}^{\;\sigma} \sF_{\rho\sigma}^{a} ), 
\ee
with $R$ being the Ricci curvature scalar. 

The symmetric tensor $\mT_{(\mu\nu)}$ and antisymmetric tensor $\mT_{[\mu\nu]}$ are derived from the current $\hmJ_{a}^{\mu}$ in Eq.(\ref{CC1}),
\be
 \mT_{\mu\nu}  & \equiv & \chih  \chi_{\mu}^{\; a} \hmJ_{a}^{\; \rho} \chi_{\rho\nu}  \equiv \mT_{(\mu\nu)} + \mT_{[\mu\nu]}, \nn \\
\mT_{(\mu\nu)}  & \equiv &  \frac{1}{2} (\mT_{\mu\nu} + \mT_{\nu\mu} ), \nn \\
 \mT_{[\mu\nu]}  & \equiv  & \frac{1}{2} (\mT_{\mu\nu} - \mT_{\nu\mu} ) .
\ee 
The tensors $\cG_{(\mu\nu)}$ and $\cG_{[\mu\nu]}$ are obtained from the current $\hcG_{a}^{\;\mu}$ in Eq.(\ref{CC}) as follows:
\be
\cG_{\mu\nu}  &  \equiv & \frac{1}{2} \chih  \chi_{\mu}^{\; a} \hcG_{a}^{\; \rho}  \chi_{\rho\nu} \equiv \cG_{(\mu\nu)} + \cG_{[\mu\nu]}, \nn \\
\cG_{(\mu\nu)} & \equiv & \frac{1}{2} (\cG_{\mu\nu}  + \cG_{\nu\mu}  ) , \; \cG_{[\mu\nu]} \equiv  \frac{1}{2} (\cG_{\mu\nu}  - \cG_{\nu\mu}  ).
\ee

\end{document}